\documentclass[aps,prd,,eqsecnum,twocolumn]{revtex4}
\usepackage{graphicx}
\usepackage{lscape}
\usepackage{indentfirst}
\usepackage{latexsym}
\usepackage{multirow}
\usepackage{tabls}
\usepackage{epsfig}

\usepackage{color}
\usepackage{amssymb}

\usepackage{amsfonts}
\usepackage{amsmath}
\usepackage{bm}
\usepackage{mathrsfs}

\def\cE{ {\cal E} }

\def\bL{ {\bf L} }

\def\bS{ {\bf S} }

\def\hatS{ {\hat{S}} }

\def\DoL{ {{\Delta}_{1\text{L}}} }
\def\DtL{ {{\Delta}_{2\text{L}}} }
\def\Dot{ {{\Delta}_{12}} }
\def\DoLtL{ {{\Delta}_{1\text{L}2\text{L}}} }
\def\DoLoL{ {{\Delta}_{1\text{L}1\text{L}}} }
\def\DtLtL{ {{\Delta}_{2\text{L}2\text{L}}} }

\def\bhS{ \hat{\bf S}}

\def\bSeff{{\bf S}_{\text{eff}}}

\def\bhL{ \hat{\bf L}}

\def\lsim{\mathrel{\rlap{\lower3pt\hbox{\hskip1pt$\sim$}}
    \raise1pt\hbox{$<$}}}              
\def\gsim{\mathrel{\rlap{\lower3pt\hbox{\hskip1pt$\sim$}}
    \raise1pt\hbox{$>$}}}         
\def\coordeq{ \, \mathrel{ \rlap{\hbox{\hskip-2.5pt$=$} }
    \raise4pt\hbox{$\cdot$}} \, } 
\def\ceq{ \mathrel{\mathop:}=}
\begin{document}

\title{Statistical constraints on binary black hole inspiral dynamics}

\author{Chad R. Galley, Frank Herrmann, John Silberholz, Manuel Tiglio} 
\affiliation{Department of Physics, Center for Fundamental Physics, Center for Scientific Computation and Mathematical Modeling, Joint 
Space Institute. University of Maryland, College Park, MD 20742, USA.}

\author{Gustavo Guerberoff}
\affiliation{Facultad de Ingenier{\'i}a, Instituto de Matem{\'a}tica y Estad{\'i}stica ``Prof.~Ing.~Rafael Laguardia''.~Universidad de la 
Rep{\'u}blica, Montevideo, Uruguay.}

\begin{abstract}
We perform a statistical analysis of the binary black hole problem in the post-Newtonian approximation by systematically sampling and evolving the parameter space of initial configurations for quasi-circular inspirals. 
Through a principal component analysis of spin and orbital angular momentum 
variables we systematically look for uncorrelated quantities and find three of them which are highly conserved in a statistical sense, both 
as functions of time and with respect to variations in initial spin orientations.  
We also look for and find the variables that account for the largest variations in the problem.  
We present binary black hole simulations of the full Einstein equations analyzing to what extent these results might carry over to the full theory in the inspiral and merger regimes. 
Among other applications these results should be useful both in semi-analytical and numerical building of templates of gravitational waves for gravitational wave detectors. 
 
\end{abstract}

\maketitle

\section{Introduction}

The inspiral and merger of binary black hole systems is of great importance in many astrophysical settings, from the large-scale structure of the universe via galactic merger trees to earth-based gravitational wave detectors (LIGO~\cite{Abbott:2007kva}, Virgo~\cite{Acernese:2008zza}, and GEO600~\cite{Willke:2007zz}) wherein stellar-mass black hole inspirals and intermediate-mass black hole mergers are the best source candidates in their frequency band. Because they encounter a small signal-to-noise ratio, these detectors rely on banks of templates of gravitational waves for their detection, which vary based on the properties of their sources. 
Within General Relativity (GR) the parameter space of initial configurations for binary black holes in quasi-circular orbit is eight-dimensional. From the no-hair theorem each black hole is uniquely characterized by its mass $m_i$, spin orientation $\hat{\mathbf{S}}_i$ and dimensionless spin magnitude $\chi_i$ ($i=1,2$) defined by spin vector $\mathbf{S}_i=\chi_i m_i^2
\hat{\mathbf{S}}_i$.  In the absence of matter there is no preferred scale in GR so that the parameter space can actually be trivially reduced to at least seven dimensions. 
Still, the cost of generating templates for a multi-dimensional parameter space through 
full numerical simulations is prohibitively expensive. Semi-analytical models are a very promising approach but they also require calibration through numerical simulations, which are still computationally expensive even if to a lesser extent. Any possible simplification or guidance in building or computing gravitational wave templates is therefore crucial. 

The post-Newtonian (PN) approximation to General Relativity is a good description of the inspiral dynamics until the black holes get close to each other~\cite{Boyle:2007ft,Hannam:2007wf}. 
Because the PN approximation is dramatically simpler than full GR --- and correspondingly much more computationally inexpensive --- it is feasible to thoroughly study the parameter space in this 
approximation numerically and obtain information that is relevant in the fully relativistic case. Among recent developments this includes statistical predictions for recoil velocities after 
a black hole collision \cite{Lousto:2009ka}, how large recoils can be suppressed due to spin alignment with the orbital angular momentum \cite{Kesden:2010ji}, whether initially uniform spin orientation distributions remain uniform during evolution, how initial and final spin orientations correlate over time, and the use of graphics processing units to accelerate these simulations~\cite{Herrmann:2009mr}. 

In this paper we present an analytical and numerical statistical analysis of the binary black hole parameter space in an initially quasi-circular orbit in the PN approximation. We numerically sample and evolve the entire $7$-dimensional space of initial configurations in the PN approximation. We focus on finding the most and least relevant variables in the spin dynamics, namely, those that account for most of the dynamics and those that are largely conserved, even in the presence of radiative corrections in the binary's evolution. This is precisely the area of Principal Component Analysis (PCA), a rather standard technique in multivariate statistical analysis that does not yet seem to have been applied to the binary black hole problem. 
To obtain approximate expressions in closed form we also implement a PCA in what we call the instantaneous approximation and validate to what extent the numerical results can be described via this analysis. Finally, we present some preliminary studies, through numerical simulations of colliding binary black holes using the full Einstein equations, aimed at studying to what extent the results of this paper carry over to the fully relativistic case 
in the inspiral to merger regimes.

%%%%%%%%%%%%%%%%%%%%%%%%%%%
\section{Post-Newtonian approximation} \label{sec:PN}
%%%%%%%%%%%%%%%%%%%%%%%%%%%

We work with the PN equations from
Ref.~\cite{Buonanno:2002fy, Buonanno:2002erratum},
which describe a quasi-circular inspiral of two spinning black holes up to 3.5PN order
in the angular frequency $\omega$ and spin effects up to 2PN order
with the covariant spin supplementary condition. The evolution is
given by a system of coupled ordinary differential equations (ODEs) for the orbital frequency $\omega$,
the individual spin vectors $\mathbf{S}_j$ of each black hole and the unit
orbital angular momentum vector $\hat{\mathbf{L}}$. For completeness, these equations are given explicitly in Appendix A. The spin vectors $\mathbf{S}_j$ are given in terms of their magnitude and orientation by $\mathbf{S}_j = m_j^2 \chi_j  \hat{\mathbf{S}}_j$ where $\chi_j \in [0,1]$ is the dimensionless Kerr spin parameter of the $j^{\rm th}$ black hole. 

We numerically evolve the system of coupled ODEs for $\omega$, $\hat{\mathbf{L}}$ and $\mathbf{S}_i$ from some 
initial frequency $\omega_i$ to a final one $\omega_f$ and systematically sample the range of masses and spin parameters $m_j, \chi_j$. We typically choose $\omega_i$ corresponding to an
initial separation of $r\approx 40 M$ and a final frequency of up to $\omega_f=0.05$ (with that exact final frequency unless otherwise 
stated), which is a
conservative estimate of where the PN equations still hold~\cite{Boyle:2007ft,Hannam:2007wf}. See \cite{Herrmann:2009mr} for details on our numerical implementation of these equations. 

In the PN approximation the free parameters of the problem have different hierarchies~\cite{Buonanno:2002fy,Buonanno:2002erratum} because the spin orientations ($\bhS_1$, $\bhS_2$) evolve as a function of time while their masses and spin magnitudes  ($m_j, \chi_j$) are constant. For this reason it is appropriate to consider 
the spin orientations as stochastic variables for each $m_j, \chi_j$~\cite{Herrmann:2009mr}. Since there is no preferred scale in 
vacuum GR, in all of our numerical simulations we fix the total mass to $M=m_1+m_2=1$. Also, we typically use, unless otherwise stated, $40,\!000$ random spin orientations to capture patterns in evolutions for each tuple $(m_1,\chi_1,m_2,\chi_2)$. We have found that this is sufficient to capture the main features of our results.

%%%%%%%%%%%%%%%%%%%%%%%%%%%%%%
\section{Principal Component Analysis}
%%%%%%%%%%%%%%%%%%%%%%%%%%%%%%

In PCA one seeks to determine the variables that are statistically relevant and to dimensionally reduce from the problem those that are not. The covariance between two stochastic variables $X,Y$ is given by 
\begin{align}
	{\rm Cov} (X,Y) & = \big\langle ( X - \langle X \rangle ) ( Y - \langle Y \rangle ) \big\rangle \nonumber \\ 
		& = \langle X Y \rangle - \langle X \rangle \langle Y \rangle
			\label{covmx_1}
\end{align}	
and provides a measure of the degree to which their fluctuations are correlated. A smaller (larger) covariance implies lower (higher) correlation. In particular, the covariance of a variable with itself is its variance (i.e., the standard deviation squared) and measures deviations from the mean value. The brackets above represent expectation values. In the context of this paper they represent averages over the unit sphere
\begin{equation}
	\langle \cdots \rangle = \frac{1}{4\pi} \int _{S^2} ( \cdots ) \, d\Omega \, , \label{sphere}
\end{equation}
i.e., expectation values with respect to the two black hole spin orientations. 

When there are multiple  stochastic variables $X_i$ ($i=1, \ldots, n$) one can construct their associated covariance matrix $\bf{C}$ with components $C_{ij} = {\rm Cov} (X_i,X_j)$. This matrix is symmetric, non-negative definite and can be diagonalized with an orthogonal transformation. If we 
denote the components of the $i^{\rm th}$ normalized eigenvector $\hat{{\bf V}}_i$ by $\hat{V}_{i}^j$ then the {\it principal components} (PCs) are the 
associated eigenmodes, 
\begin{equation}
	\cE_i  = \hat{V}_{i}^1X_1 + \ldots + \hat{V}_{i}^nX_n \, .
\end{equation}
The PCs are {\it uncorrelated}, a consequence of the orthogonality of the eigenvectors, and their associated eigenvalues $\lambda$ are their variances,
\begin{equation}
	{\rm Cov}(\cE_i, \cE_j ) = \lambda_i \delta_{ij} \, .
\end{equation}
The fact that, by construction, principal components are uncorrelated with each other is important since they provide independent pieces of statistical information. 

The smaller an eigenvalue $\lambda_i$ the more likely that the corresponding linear combination $\cE_i$ will not deviate from its average value for a randomly chosen pair of spin orientations.
Therefore, if there exist small eigenvalues then the associated PCs are largely {\it conserved} in a statistical sense. Conversely, the larger an eigenvalue is then the more relevant the associated PC is in describing the dynamics and variations in the problem.

There are two {\em related but different} senses in which a principal component with small variance will be shown to be semi-conserved in this paper. The first is 
in the usual sense, i.e., being constant as a function of time for an arbitrary but {\em fixed} initial 
spin configuration. The second is in the statistical sense that deviations of a principal component from the mean value are small for arbitrary but fixed initial and final times over a set of runs with the same masses and spin magnitudes but random spin orientations. In other words, the principal component is essentially constant with respect to sampling {\em spin orientations}. A small variance automatically implies approximate 
conservation in the second sense but not necessarily in the first one.

We emphasize that the interest here is not only in those PCs that have the smallest variances (and thus identify semi-conserved quantities) but also in those with the largest variances, which encode the most information about the inspiral dynamics.

%%%%%%%%%%%%%%%%%%%%%
\section{Variables for PCA}
%%%%%%%%%%%%%%%%%%%%%
In this paper we focus on analyzing the statistical properties of differences between final and initial values of the following scalar products between the unit orbital angular momentum and spin vectors 
in the following combinations: 
\begin{widetext}
\begin{eqnarray}
	\Delta (\bhS_1 \cdot \bhL)   &=& \bhS_1 \cdot \bhL |_f - \bhS_1 \cdot \bhL |_i \label{delta1L}  \ceq \DoL \\
	\Delta (\bhS_2 \cdot \bhL)  & = & \bhS_2 \cdot \bhL |_f - \bhS_2 \cdot \bhL |_i  \ceq \DtL  \label{delta2L} \\
	\Delta (\bhS_1 \cdot \bhS_2) &=& \bhS_1 \cdot \bhS_2 |_f - \bhS_1 \cdot \bhS_2 |_i \label{delta12} \ceq \Dot \\
\Delta [ (\bhS_1 \cdot \bhL) (\bhS_2 \cdot \bhL) ] & = & (\bhS_1 \cdot \bhL) |_f (\bhS_2 \cdot \bhL) |_f  - 
	 (\bhS_1 \cdot \bhL) |_i (\bhS_2 \cdot \bhL) |_i   \ceq \DoLtL   \label{delta1L2L} \\
\Delta [ (\bhS_1 \cdot \bhL)^2 ] & = & (\bhS_1 \cdot \bhL)^2 |_f   - (\bhS_1 \cdot \bhL)^2 |_i  \ceq \DoLoL \label{delta1L1L} \\
	\Delta [ (\bhS_2 \cdot \bhL)^2 ]  & = & (\bhS_2 \cdot \bhL)^2 |_f  - (\bhS_2 \cdot \bhL)^2 |_i   \ceq \DtLtL  \, . \label{delta2L2L} 
\end{eqnarray}
\end{widetext}
Other combinations are possible and might actually provide further insight. 

In our numerical simulations and analytical calculations we choose the initial spin orientations with uniform and uncorrelated probability distributions. The above scalar products are then also initially uncorrelated and their expectation values of (\ref{delta1L})-(\ref{delta1L2L}) vanish,
$$
\langle \bhS_1 \cdot \bhS_2 \rangle|_i= \langle \bhS_1 \cdot \bhL \rangle|_i = 
\langle \bhS_2 \cdot \bhL \rangle |_i = 0
$$
while those of (\ref{delta1L1L}) and (\ref{delta2L2L}) are
$$
\langle ( \bhS_1 \cdot \bhL )^2 \rangle |_i = \langle ( \bhS_2 \cdot \bhL )^2 \rangle |_i = 1/3
$$

The orbital angular momentum and spin orientations naturally become correlated due to spin-orbit and 
spin-spin interactions as each of these binary black hole configurations evolve in time. However, at least within the PN approximation here considered, the orbital angular momentum and spin vectors remain perfectly uniformly distributed~\cite{Herrmann:2009mr}. For example, a Kolmogorov-Smirnov test for a representative  configuration returns a p-value of $\sim 10^{-5}$ when testing for lack of uniformness~\cite{Herrmann:2009mr}. Higher PN expansions might introduce small biases~\cite{Lousto:2009ka} but if so they appear to be at a level in which approximating the mean of the above scalar products at any instant of time by zero is a very good approximation. 

%%%%%%%%%%%%%%%%%%%%%%%
\section{Instantaneous approximation}\label{sec:IA}
%%%%%%%%%%%%%%%%%%%%%%%%%%%%%
Computing the covariance matrix requires sampling a large part of the space of initial spin orientations for each fixed set of binary black hole system parameters (tuples of masses and spin magnitudes) and necessitates computing the solutions of the PN equations repeatedly for each set. Over long time periods (large $t_f - t_i$) we solve these equations numerically but we can also gain some insight into the structure of principal components by studying the evolution of the system over short durations.

The changes in the scalar products used in our analysis can be calculated in the approximation where $t_f = t_i + \Delta t$ for a small interval of time $\Delta t$. 
 Using the equations of motion for the spin and orbital angular momentum and keeping terms through $O(\Delta t)$ gives
\begin{align}
		\DoL \approx {} & \frac{ 3 m_2 \chi_2 \omega^2 \Delta t}{ 2 M } \bhL \cdot (\bhS_1 \times \bhS_2 ) \nonumber \\
		& \times \bigg( M -  (M \omega)^{1/3} m_1 \chi_1 \bhS_1 \cdot \bhL \bigg) 
			\label{S1L_2} \\
	\DtL = {} & \DoL {\rm ~with~} 1 \leftrightarrow 2 
			\label{S2L_2}  \\
	\Dot \approx {} &  \frac{3 \omega^{5/3} \Delta t}{ 2 M^{4/3} } \bhL \cdot (\bhS_1 \times \bhS_2 ) \bigg( - m_1^2 + m_2^2 \nonumber \\
		& + (M\omega)^{1/3} (m_1^2 \chi_1 \bhS_1 - m_2^2 \chi_2 \bhS_2 ) \cdot \bhL \bigg)  
			\label{S1S2_2} 	\\
	\DoLtL \approx {} & - \frac{3 \omega^2 \Delta t }{2} \, \bhL \cdot (\bhS_1 \times \bhS_2) \nonumber \\		& \times  \bigg(m_1 \chi_1 \bhS_1 \cdot \bhL - m_2 \chi_2 \bhS_2 \cdot \bhL \bigg)
			\label{S1LS2L_2} \\
	\DoLoL \approx {} & \frac{3 m_2 \chi_2 \omega^2 \Delta t }{ M } (\bhS_1 \cdot \bhL)  \, [ \bhL \cdot ( \bhS_1 \times \bhS_2 ) ] \nonumber \\
		& \times \bigg( M - m_1 \chi_1 (M \omega)^{1/3} \bhS_1 \cdot \bhL \bigg) 
			\label{S1LS1L_2} \\
	\DtLtL = {} & - \DoLoL {\rm ~with~} 1 \leftrightarrow 2 
			\label{S2LS2L_2}
\end{align}
where the right sides are evaluated at the initial time (frequency).

In calculating the covariance of the scalar products we must first determine the expectation values of (\ref{S1L_2})-(\ref{S2LS2L_2}) through $O(\Delta t)$. Doing this exactly is challenging because $\Delta t$ is found by solving the equations of motion for $\omega (t)$ and depend on the spin scalar products, which are stochastic variables,
\begin{equation}
	\Delta t = \Delta t (\omega, \bhS_1, \bhS_2 , \bhL )  .
\end{equation}
We perform this calculation because we use a common final frequency (as opposed to a common final time) in our analysis to provide a less gauge-dependent stopping criteria 
dependence. Physically, the interactions between the spins can cause a slight repulsion or attraction between the masses that increases or decreases, respectively, the elapsed time of the inspiral from $\omega_i$ to $\omega_f$. Therefore, we cannot simply factor $\Delta t$ outside of the average  nor are we able to solve the equations of motion analytically to find an expression for $\Delta t$. However, the spins affect the orbital frequency beginning at 1.5PN order and are small perturbations. Therefore, we can assume $\Delta t$ is independent of spin to a good approximation, and so the expectation values of the scalar products in (\ref{S1L_2})-(\ref{S2LS2L_2}) all vanish in this approximation since the spins are uncorrelated and have zero mean at the initial time.

The elements of the covariance matrix are then calculated by forming the appropriate covariances of (\ref{S1L_2})-(\ref{S2LS2L_2}), which are all $O(\Delta t^2)$ to leading order. Since the spin directions are uniformly distributed over a 2-sphere then the expectation value of an odd number of spins vanishes and
\begin{equation}
	\langle \hatS_k^{i_1} \cdots \hatS_k^{i_{2n} } \rangle = \frac{1}{(2n+1)!!} ( \delta^{i_1 i_2} \cdots \delta ^{i_{2n-1} i_{2n} } + \cdots )
\end{equation}
for an even number of spins where the last $\cdots$ indicates all possible pairings of indices and $k=1,2$. After some algebra we find that the elements of the covariance matrix are
\begin{equation}
	C_{ij} = (\Delta t)^2 \tilde{C}_{ij} + O(\Delta t^3)
\end{equation}
where the $\tilde{C}_{ij}$ are in some cases too lengthy to display in full detail. 
Since $C_{ij} = O(\Delta t^2)$ at leading order then so also are the eigenvalues so that
\begin{align}
	 \tilde{{\bf C}} {\bf V}_{j} =  \tilde{\lambda}_j {\bf V}_{j}
\end{align}
where $\lambda = (\Delta t)^2 \tilde{\lambda}$. We order the eigenvectors $\bf{V}_i$ and associated $\lambda_i$ such that $\lambda_1 \geq \lambda_2 \geq ... \geq \lambda_n$ for an $n$ x $n$ covariance matrix.

%%%%%%%%%%%%%%%%%%%%%%%%%%%%%%
\section{Spin-orbit variables ({\bf SO})}  \label{sec:SO}
%%%%%%%%%%%%%%%%%%%%%%%%%%%%%%

We illustrate this approach in detail with a simple case and make contact with previous conservation results. We start building towards our more general result by first doing a PCA using only the two 
spin-orbit variables in (\ref{delta1L}) and (\ref{delta2L}), 
\begin{equation}
\DoL \ceq \Delta (\bhS_1 \cdot \bhL)  \;\;\; , \;\;\; \DtL \ceq \Delta (\bhS_2 \cdot \bhL) \, .  \label{SOPCA}
\end{equation}
We remind the reader that we include spin-spin interactions in our simulations and analytical calculations by using the PN equations of motion 
described in Section~\ref{sec:PN} and the Appendix, but in this section we only 
use the variables in (\ref{SOPCA}) for the PCA. 

For each black hole mass and spin magnitudes ($m_j,\chi_j$) the covariance matrix for the variables (\ref{SOPCA}) is
\begin{equation}
C  = 
\left(
\begin{array}{cc}
{\rm Cov} ( \DoL, \DoL ) &  {\rm Cov} ( \DoL, \DtL ) \\
{\rm Cov} ( \DtL, \DoL ) &  {\rm Cov} ( \DtL, \DtL ) 
\end{array}
\right)  \, , 
\label{covmxSO}
\end{equation}
where the entries can come either from our numerical simulations or the instantaneous approximation. 
We then diagonalize $C$ to find the principal components.

\subsection{Numerical simulations}

From our numerical simulations we find that, sampling across many random initial spin orientations, each of the principal components has zero mean over time (to numerical accuracy), $\langle \Delta \cE_j^{\text{SO}} \rangle = 0$, a consequence of the spin orientation distributions remaining highly uniform during the inspiral.

We find $\lambda_2$ to be in the range $\sim 10^{-9}$ to $10^{-4}$ for the parameters we sampled ($m_1 \in [0.1,0.9]$ and $\chi_{1,2} \in [0.1, 1]$). Furthermore, $\lambda_2$ grows with both spin magnitudes, which is expected, but {\em increases} as the equal mass case is approached (we explain the reason for this behavior below); see Fig.~\ref{fig:m1_lam}. 

\begin{figure}
\includegraphics[width=\columnwidth]{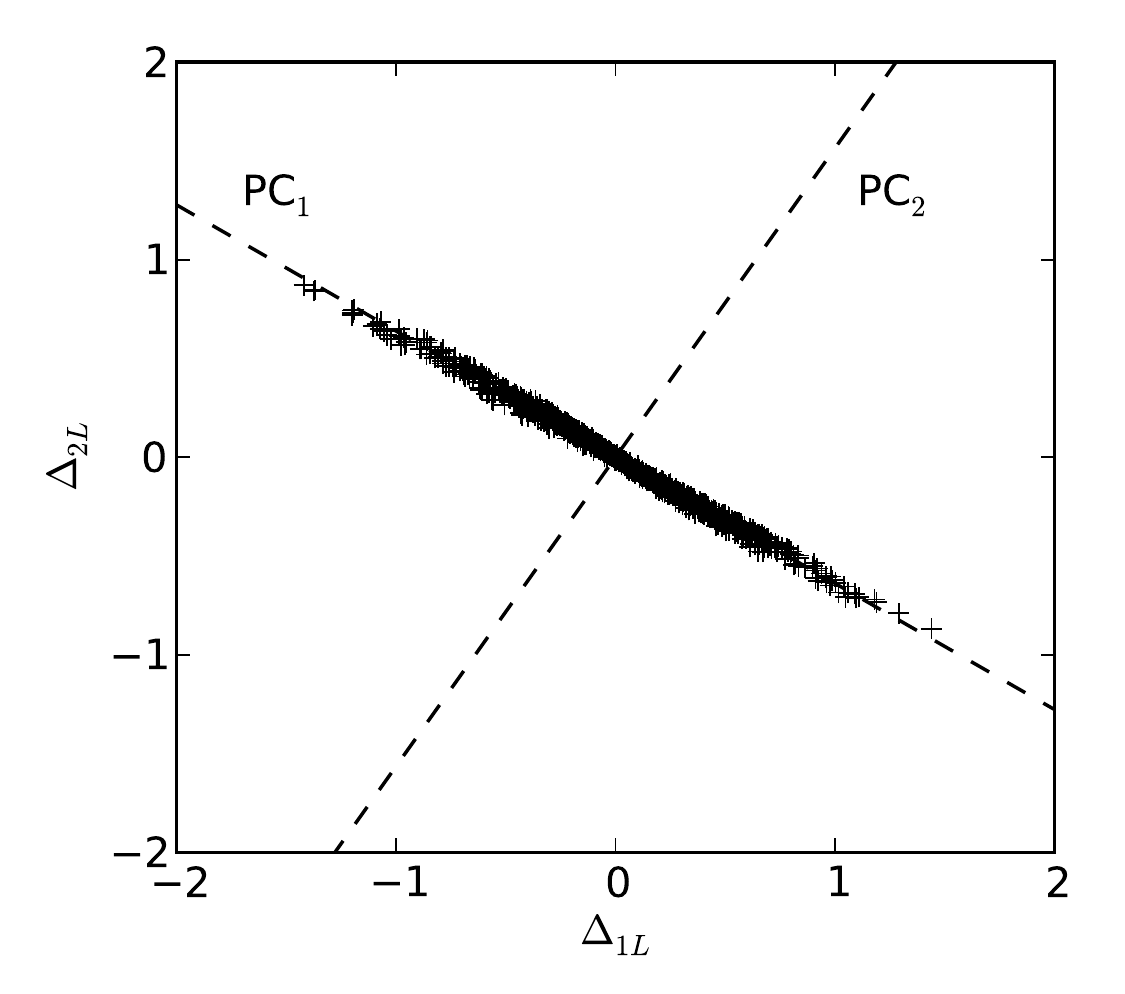}
\caption{A graphical representation of the principal components for the spin-orbit variables of Section \ref{sec:SO} and the numerical data of $\DoL$ and $\DtL$ for a binary black hole system with $m_1 = 0.4$ and maximal spin magnitudes. We show only $10^3$ points here but the PCs were computed with $10^5$ random spin orientations.}
\label{fig:PC_2d}
\end{figure}

As an example with $(m_1,m_2, \chi_1,\chi_2) = (0.4, 0.6, 1.0, 1.0)$, Figure \ref{fig:PC_2d} shows a graphical representation of the principal components overlaid on a scatter plot of the $\DoL$ and $\DtL$ data from $1,\!000$ of our $100,\!000$ numerical simulations using random initial spin orientations. Notice that the first PC, which points along the direction of the eigenvector $\hat{\bf{V}}_1$ with the largest eigenvalue $\lambda_1$, captures the largest variation in the data while the second PC, pointing along $\hat{\bf{V}}_2$, indicates that there is very little spread in the data in that direction, which is also implied by the smallness of $\lambda_2$ relative to $\lambda_1$.

The coefficients of the PCs, which are just the components of the corresponding eigenvectors, are functions of the black hole masses and spin magnitudes but also of their initial and final orbital frequencies,  
$$
\hat{{\bf V}}_{j} = \hat{{\bf V}}_{j}(m_1,\chi_1,\chi_2, \omega_i, \omega_f) ~ , ~~ j =1,2 \, .
$$
However, we have numerically found that the dependence on the initial and final frequencies is rather weak. This, together with the fact that $\Delta \cE_2$ has zero mean, implies that 
the function
$$
\cE_2^{\text{SO}} = \hat{V}_{2}^1 \, \bhS_1 \cdot \bhL  + \hat{V}_{2}^2 \, \bhS_2 \cdot \bhL 
$$
is not only constrained to be nearly constant, in a statistical sense, for all spin orientations for any black hole masses and spin magnitudes but also to be approximately constant as a function of time. This motivates and justifies our instantaneous approximation described in Section~\ref{sec:IA}.

\subsection{Instantaneous approximation}

In the instantaneous approximation we use the expressions in (\ref{S1L_2}) and (\ref{S2L_2}) to calculate the covariance matrix in (\ref{covmxSO}) to leading order in $\Delta t$ to find that $C_{ij} = (\Delta t)^2 \tilde{C}_{ij} + O(\Delta t^3)$ with
\begin{align}
	\tilde{C}_{11} = {} & \frac{ m_2^2 \chi_2^2 \omega^4 }{ 10 M^2 } \big( 5 M^2 + m_1^2 \chi_1^2 (M\omega)^{2/3} \big) \\
	\tilde{C}_{12} = {} & - \frac{1}{2} m_1 m_2 \chi_1 \chi_2 \omega^4= \tilde{C}_{21} \\
	\tilde{C}_{22} = {} & \tilde{C}_{11} {\rm ~with~} 1 \leftrightarrow 2 .
\end{align}
We then diagonalize $\tilde{C}$ to find the PCs as with our numerical simulations. We find that
\begin{eqnarray}
\Delta \cE_1^{\text{SO}} & = &  \mu_1 ( m_2 \chi_2 \DoL - m_1 \chi_1 \DtL )  \label{DE1_2d} \, ,   \\
\Delta \cE_2^{\text{SO}}  & = & \mu_2 \Delta ( \bS_0 \cdot \bhL ) =: \Delta_{\text{0L}} \label{DE2_2d} \, , 
\end{eqnarray}
with $\mu_1 = [ (m_1\chi_1)^2 + (m_2\chi_2)^2 ]^{-1/2}$ and $\mu_2 =  \mu_1/ M$. Here,
$$
\mathbf{S}_0 = \left(1+ \frac{m_2}{m_1}\right) \mathbf{S}_1 + \left(1+ \frac{m_1}{m_2} \right) \mathbf{S}_2  
$$ 
is one of the two spin vectors introduced in \cite{eob2} in the context of the effective one-body (EOB) formalism.  
In that reference it was also shown that at 2PN order in spin effects the scalar product between 
the  {\it effective} spin 
$$
\bSeff=\left(1+ \frac{3m_2}{4m_1}\right) \mathbf{S}_1 + \left(1+ \frac{3m_1}{4m_2} \right) \mathbf{S}_2 
$$
and the unit orbital angular momentum $\bSeff \cdot \bhL$ 
is a constant of motion if ignoring spin-spin interactions and radiative effects (see also \cite{Buonanno:2002fy}). That is, for each initial configuration  
this scalar product is constant in time, 
$$
\Delta_{\text{eff\,L}} \equiv \Delta ( \bSeff \cdot \bhL ) = 0 .
$$ 
More recently, it has also been shown that at the same PN order the quantity $\mathbf{S}_0 \cdot \mathbf{L}/|| \mathbf{L} ||^2$
is also a constant of motion if including quadrupole-monopole interactions, i.e., 
\begin{equation}
\Delta \big( \mathbf{S}_0 \cdot \bhL/ || \mathbf{L} ||^2 \big) = 0 \label{DS0L}  ,
\end{equation} 
and ignoring spin-spin interactions and radiation reaction \cite{Racine:2008qv}.

From (\ref{DE1_2d}) and (\ref{DE2_2d}) the instantaneous approximation implies that the PC with the smallest variance when {\it including} spin-spin and radiative effects in the equations of motion is neither $\Delta  ( \mathbf{S}_0 \cdot \bhL/ || \mathbf{L} ||^2 )$ nor $\Delta_{\text{eff\,L}}$ but $\Delta_{\text{0L}}$ instead [cf.,~Eqs.~(\ref{DE2_2d})]. The differences between $\Delta_{\text{0L}}$ and $\Delta  ( \mathbf{S}_0 \cdot \bhL/ || \mathbf{L} ||^2 )$ are detailed in Section~\ref{sec:sscomp}.

Furthermore, the instantaneous 
eigenvalues (variances) in the instantaneous approximation are given by $\lambda_j ^{\text{SO}} = (\Delta t)^2 \tilde{\lambda}_j^{\text{SO}}$ ($j=1,2$) with  
\begin{align}
	\tilde{\lambda}_1^{\text{SO}} = {}& \frac{ \omega^4 }{ 10 M^{4/3} } \big[ 5M^{4/3} (m_1^2\chi_1^2 + m_2^2\chi_2^2) \nonumber \\
	& {\hskip0.65in} +  m_1^2 \, m_2^2 \, \chi_1^2 \, \chi_2^2 \, \omega^{2/3}  \big]  
 		\label{SO_lam1}\\
	\tilde{\lambda}_2^{\text{SO}} = {} &  \frac{m_1^2 \, m_2^2 \, \chi_1^2 \, \chi_2^2 \, \omega^{14/3}}{10 M^{4/3}} 		\label{SO_lam2} \, . 
\end{align}
These expressions provide insight regarding the dependence of the variances on the black hole masses and spin magnitudes. 
For example, by analyzing Eq.~(\ref{SO_lam2}) one can see that it increases with both spin magnitudes (which is expected) but also that it {\em increases} as the equal mass case is approached, as we already found from our numerical simulations (see Fig.~\ref{fig:m1_lam}).

\subsection{Comparisons with other quantities}\label{sec:sscomp}

$\Delta \cE_1^{\text{SO}}$ is the PC describing the largest variations in the data and $\Delta \cE_2^{\text{SO}}$ is essentially conserved, as both a function of time and with respect to initial spin orientation variations,
\begin{equation}
\cE_2^{\text{SO}} =  \mu_2  \, \bS_0 \cdot \bhL  \approx \mbox{constant.} 
	\label{E2_2d}
\end{equation}
In the approximations of Ref.~\cite{Racine:2008qv}  $\Delta \cE_2^{\text{SO}}$ and (\ref{DS0L}) are essentially equivalent to each other. Here, however, we are including radiative effects and spin-spin interactions and find that $\Delta \cE_2^{\text{SO}}$ is conserved much better than $\mathbf{S}_0 \cdot \bhL/ || \mathbf{L} ||^2$.

\begin{figure}
\includegraphics[width=\columnwidth]{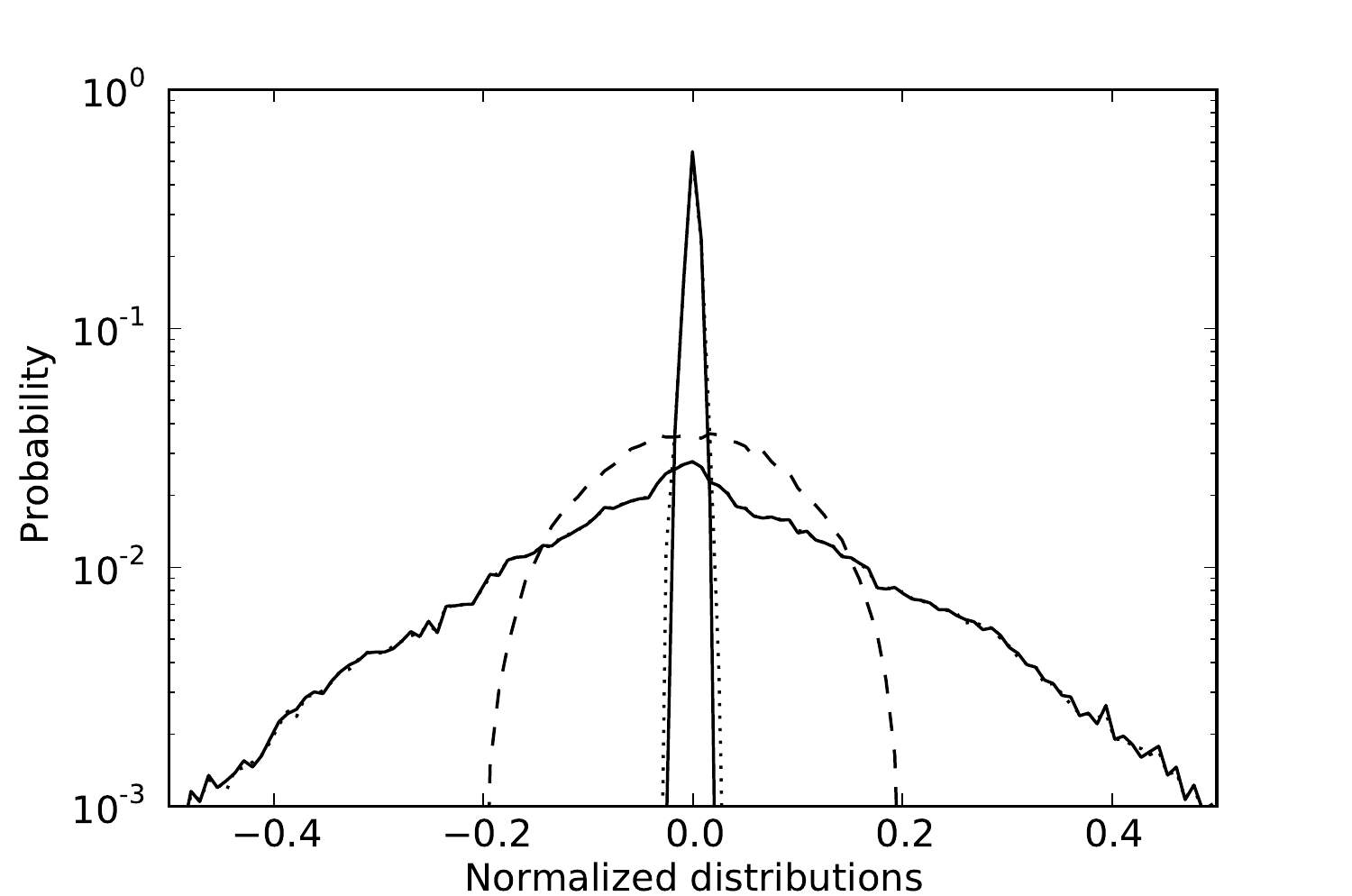}
\caption{Probability distributions for $\Delta \cE_1^{\rm SO}$ (wide) and $\Delta \cE_2 ^{\rm SO}$ (narrow) for maximally spinning black holes with $m_1 = 0.4$.
The solid lines are the numerical principal components and the dotted lines are principal components from the instantaneous approximation in (\ref{DE1_2d}) and (\ref{DE2_2d}); note that the approximations are virtually indistinguishable from the numerical results. The dashed line shows the distribution of the quantity in (\ref{DS0L}). All the distributions have been normalized with respect to their maximum possible values so that the maximum range in these plots is $[-1,1]$.  Notice that $\Delta \cE_1^{\text{SO}}$ accounts for most of the variation in the data and  $\Delta \cE_2^{\text{SO}}$ is sharply peaked around zero.}
\label{fig:2d}
\end{figure}

For example, Figure \ref{fig:2d} shows probability distributions from our numerical simulations for maximally spinning black holes with $m_1=0.4$.   Displayed 
are the two numerical PCs $\Delta \cE_1^{\text{SO}}, \Delta \cE_2^{\text{SO}}$ (solid lines), along with their instantaneous expressions 
(\ref{DE1_2d}), (\ref{DE2_2d}) (dotted lines),  as well as the quantity in expression (\ref{DS0L}) (dashed line). The numerical values and instantaneous approximations for the PCs 
are essentially indistinguishable.  
 The variances for the PCs are $\lambda_1 = 3.5\times 10^{-2}$ and $\lambda_2 =3.9 \times 10^{-5}$, respectively, while the latter is to be compared with a variance of 
$6.9 \times 10^{-2}$ for  (\ref{DS0L}). Notice that $\Delta \cE_2^{\text{SO}}$ has a considerably sharper distribution compared to  expression (\ref{DS0L}) in Figure \ref{fig:2d}. In fact, the variance of $\mathbf{S}_0 \cdot \bhL/ || \mathbf{L} ||^2$ is of the order of magnitude of the first principal component $\Delta \cE_1^{\text{SO}} $, 
which is the one describing the {\em largest} variation in the data.

In Figure \ref{fig:2d} we have not shown the probability distribution for $\Delta_{\text{eff\,L}}$ because it is practically indistinguishable from $\Delta_{\text{0L}}$. This is expected because $\bS_0$ and $\bSeff$ are defined as sums of the individual spins with similar order of magnitude coefficients (in fact, $\bS_0$ and $\bSeff$ are proportional to each 
other in the equal mass and spin case). 

%%%%%%%%%%%%%%%%%%%%%%%%%
\section{Spin-orbit and spin-spin variables (SOSS)} \label{sec:SS}
%%%%%%%%%%%%%%%%%%%%%%%%%
Section~\ref{sec:SO} ignored spin-spin variables in the analysis. Here
we extend the previous analysis using $\DoL$ and $\DtL$ by including
$\Dot$ in the PCA. There are now three principal components, which are
linear combinations of these variables, and as before we rank them by
decreasing variances.

From our numerical simulations we find the smallest variance
$\lambda_3$ has the same range as the smallest variance in Section~\ref{sec:SO}. However, the behavior is different with respect to
the mass ratio. Namely, for any pair of spin magnitudes the variance
$\lambda_3$ is smallest for small mass ratios, grows as the masses get
comparable --- taking a maximum around $m_1 \sim 0.4$ --- and then
decreases as the equal mass case is approached (see
Fig.~\ref{fig:m1_lam} for the maximally spinning case
$\chi_1=\chi_2=1$). As in Section~\ref{sec:SO}, for a fixed mass ratio
the variance is monotonically increasing with respect to both spin
magnitudes.

In the instantaneous approximation the infinitesimal covariance matrix
is ${\bf C } = (\Delta t)^2 \tilde{{\bf C}} + O(\Delta t^3)$. with
\begin{widetext}
\begin{equation}
\tilde{{\bf C}} =\frac{ x^5 }{ 10 M^6 }
\left(
\begin{array}{ccc}
	 \displaystyle\frac{ x S_2^2 }{ \eta^2 M^2}  \left(5 m_1^2 + x S_1^2 / M^2  \right)  
		& -\displaystyle\frac{ 5 x S_1 S_2 }{ \eta }
		& -\displaystyle\frac{ x^{1/2} S_2 }{ \eta} \left( 5  m_1 \delta m  +x S_1^2 / M^2 \right) \\
	 -\displaystyle\frac{5 x S_1 S_2}{\eta} 
		& \displaystyle\frac{ x S_1^2 }{ \eta^2 M^2 } \left(5 m_2^2 +  x S_2^2 / M^2 \right) 
		& \displaystyle\frac{ x^{1/2} S_1 }{ \eta } \left( 5 m_2 \delta m - x S_2^2 / M^2 \right) \\
	-\displaystyle\frac{ x^{1/2} S_2 }{ \eta} \left( 5  m_1 \delta m  +x S_1^2 / M^2 \right) 
   		& \displaystyle\frac{ x^{1/2} S_1 }{ \eta } \left( 5 m_2 \delta m - x S_2^2 / M^2 \right) 
		& 5 M^2 \delta m^2 + x ( S_1^2 + S_2^2)   
\end{array}
\right)
\end{equation}
\end{widetext}
where $S_i \equiv | {\bf S}_i | = m_i^2 \chi_i$, $\eta = m_1 m_2 / M^2$ is the symmetric mass ratio, $M = m_1 + m_2$ is the total mass, $\delta m = m_1 - m_2$ and $x = (M \omega)^{2/3}$.

The principal components and variances are straightforward to compute but their expressions are rather lengthy so we only display those corresponding 
to the smallest eigenvalue. In this case
$$
\lambda_3^{\text{SOSS}} = (\Delta t)^2 \tilde{\lambda}_3^{\text{SOSS}}
$$
with $\tilde{\lambda}_3^{\text{SOSS}}=0$, and its associated (non-normalized) eigenvector is 
\begin{align}
	{\bf V}_{3}^{\text{SOSS}} = \big( m_1^2 \chi_1, m_2^2 \chi_2 , m_1 \chi_1  m_2 \chi_2 (M \omega)^{1/3}  \big)   \, . 
\end{align}
The third principal component is thus 
\begin{align}
	\Delta \cE_3^{\text{SOSS}} =  \frac{1}{|| {\bf V}_3^{\text{SOSS}} ||}\left( \frac{(M \omega)^{1/3}}{ m_1 m_2 } \Delta ( \bS_1 \cdot \bS_2 ) + \Delta ( \bS \cdot \bhL ) \right)  \, , 
	\label{lincomb_1}
\end{align}
with 
$$
\bS = \bS_1 + \bS_2
$$ the total spin angular momentum. In the instantaneous approximation the variance of $\Delta \cE_3^{\text{SOSS}}$ is zero (as is its expectation value) and equals the sum of non-negative numbers. It therefore follows that $\Delta \cE_3^{\text{SOSS}}$ itself vanishes and that $\cE_3^{\text{SOSS}} |_f = \cE_3^{\text{SOSS}} |_i + O(\Delta t)$, or
\begin{align}
	(M \omega)^{1/3} \Delta (\bS_1 \cdot \bS_2 ) = - m_1 m_2 \Delta ( {\bf S } \cdot \bhL ) + O( \Delta t) .
\end{align}
Similarly, in the extreme mass ratio limit, $m_1 \ll m_2$, (\ref{lincomb_1}) becomes
\begin{align}
	\Delta \cE_3^{\text{SOSS}} =  \frac{m_1}{m_2} \chi_1 ( m_2 \omega)^{1/3} \Delta ( \bhS_1 \cdot \bhS_2) + \Delta ( \bhS_2 \cdot \bhL ) + O(\Delta t)
	\nonumber
\end{align}
to leading order in the mass ratio. 
The fact that in our numerical simulations we find $\Delta \cE_3 $ to be nearly conserved for finite times implies that   
\begin{align}
	(M \omega)^{1/3} \Delta (\bS_1 \cdot \bS_2 ) \approx - m_1 m_2 \Delta ( {\bf S } \cdot \bhL ).
\end{align}

%%%%%%%%%%%%%%%%%%%%%%%%%
\section{Spin-orbit and spin-spin variables -- Nonlinear (NL)}
%%%%%%%%%%%%%%%%%%%%%%%%%

Here we consider the six variables from Eqs.~(\ref{delta1L})-(\ref{delta2L2L}), which include nonlinear dependence on the spin-orbit variables.  
 When performing a PCA in these variables we find that there are {\em three} principal components ($\Delta \cE_4^\text{NL}$, $\Delta \cE_5^\text{NL}$ and $\Delta \cE_6^\text{NL}$ in order of decreasing variance) with small variances. From our numerical simulations we find that these eigenvalues are all bounded from above by small numbers for any binary black hole mass and spin magnitudes. More precisely, for the span of frequencies considered in this paper they are in the ranges $\lambda_4^\text{NL} \sim 10^{-7}$ to $10^{-2}$, $\lambda_5^\text{NL} \sim 10^{-9}$ to $10^{-4}$ and $\lambda_6^\text{NL} \sim 10^{-11}$ to $10^{-5}$.

\begin{figure}
\includegraphics[width=\columnwidth]{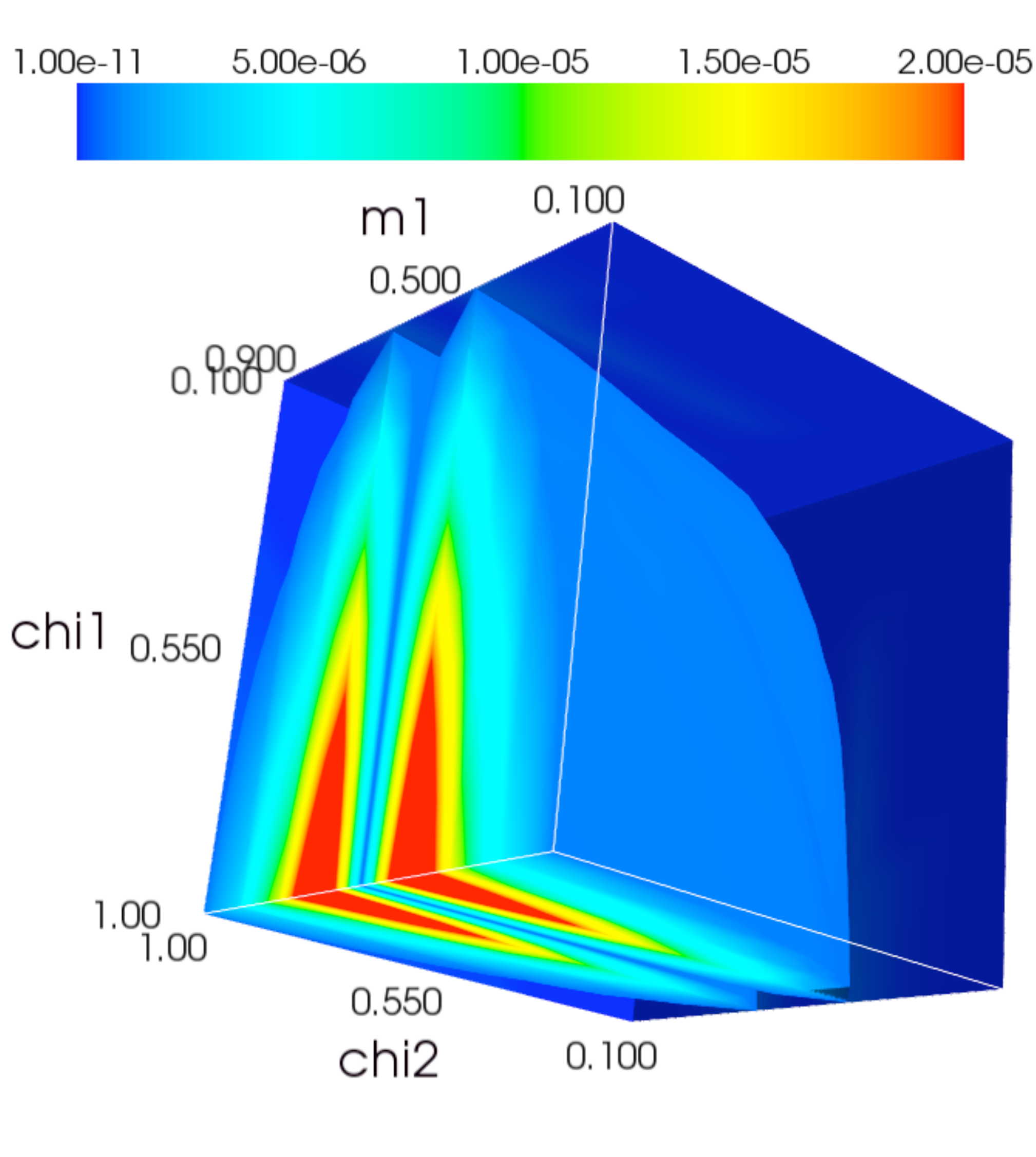}
\caption{Variance of the principal component $\Delta \cE_6^\text{NL}$ as a function of the black hole spin magnitudes $\chi_1,\chi_2$ and 
their masses (recall that throughout this paper the total mass is set to $m_1+m_2=1$ in all numerical simulations). }
\label{fig:lam6_6d}
\end{figure}

As an example, Figure \ref{fig:lam6_6d} shows the variance of $\Delta \cE_6^\text{NL}$ (i.e., the eigenvalue $\lambda_6^\text{NL}$) as a function of the black hole masses and spin magnitudes for evolutions up to a final frequency of $\omega_f=0.05$. For any pair of spin magnitudes the variance is smallest for very different binary masses, grows as the latter get comparable, and then rapidly decreases to zero as the equal mass case is approached.  For a fixed mass ratio the variance is monotonically increasing with respect to both spin magnitudes. From Fig.~\ref{fig:lam6_6d} one can also notice that we can place the bound $\lambda_6^\text{NL} \lsim 2 \times 10^{-5}$ for any possible configuration of binary black hole masses and spin magnitudes.

We emphasize that the PCs $\cE_4^\text{NL}$, $\cE_5^\text{NL}$ and $\cE_6^\text{NL}$ are distinctly new quantities different from either $\bS_0 \cdot \bhL$, $\bSeff \cdot \bhL$ or $\bS_0 \cdot \bL / || \bL || ^2$. This can be noticed, for example, from 
Table \ref{tab:PCAuneqmass} where we explicitly list the components of $\cE_4^\text{NL}$, $\cE_5^\text{NL}$ and $\cE_6^\text{NL}$ for a 
particular mass and spin magnitude configuration. One can see that the spin-spin and the non-linear spin-orbit contributions are non-negligible. In addition, recall that the principal components are by construction statistically uncorrelated. 
\begin{table}
	\caption{Components of the normalized eigenvectors for the principal components with the three smallest variances for a binary black hole system with $m_1 = 0.4$ ($m_1+m_2=1$) and maximal spin magnitudes. }
\begin{tabular}{| c | c | c | c |}
	\hline
	~ & $\hat{V}_4$ & $\hat{V}_5$ & $\hat{V}_6$ \\
	\hline
	$\lambda$ & $1.97 \times 10^{-2}$ & $1.05 \times 10^{-4}$ & $3.27 \times 10^{-5}$ \\
	\hline
	$\Delta_{1L}$ & $0.625$ & $0.051$ & $0.538$ \\
	$\Delta_{2L}$ & $-0.408$ & $0.121$ & $0.832$ \\
	$\Delta_{12}$ & $-0.604$ & $0.012$ & $0.014$ \\
	$\Delta_{1L2L}$ & $-0.160$ & $0.757$ & $-0.132$ \\
	$\Delta_{1L1L}$ & $0.054$ & $0.248$ & $-0.018$ \\
	$\Delta_{2L2L}$ & $0.224$ & $ 0.590$ & $ -0.040$ \\
	\hline
\end{tabular}
\label{tab:PCAuneqmass}
\end{table}

\begin{figure}
\includegraphics[width=\columnwidth]{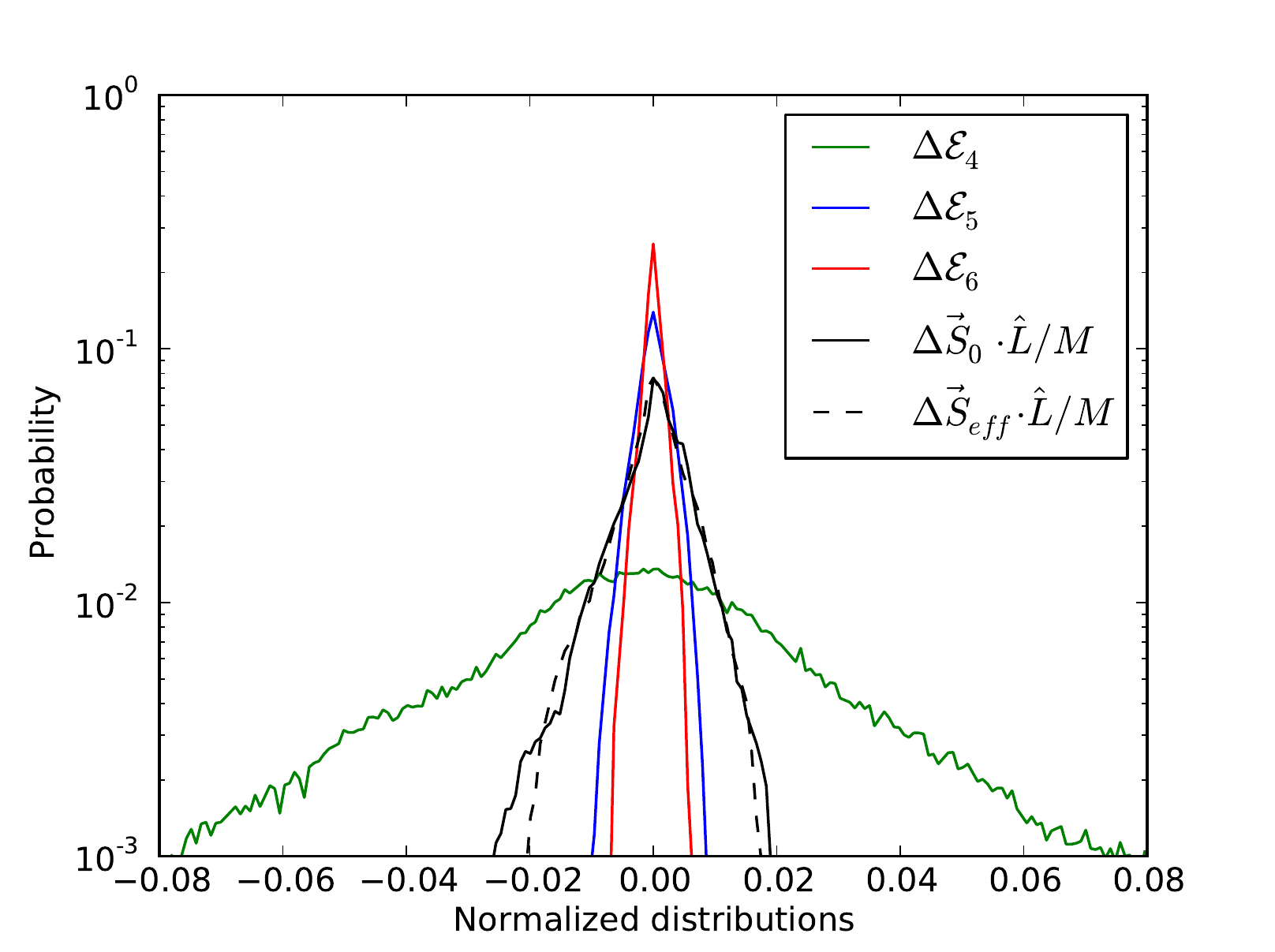}
\caption{Probability distribution of the three principal components with the smallest variance ($\Delta \cE_4^\text{NL}$ (green), $\Delta \cE_5^\text{NL}$ (blue) and $\Delta \cE_6^\text{NL}$ (red)). For comparison, $\Delta _{0\text{L}}$ (solid black) and $\Delta _{\text{eff L}}$ (dashed), which are truly zero when ignoring spin-spin and radiative corrections, are also plotted.}
\label{fig:norm_dist6d}
\end{figure}

The PCs $\cE_4^\text{NL}$, $\cE_5^\text{NL}$ and $\cE_6^\text{NL}$ are semi-conserved quantities in the two senses used in this paper in the presence of both {\it radiative corrections} (in the form of radiation reaction driving the inspiral) and {\it spin-spin interactions}. This is in contrast to $\bSeff \cdot \bhL$ and $\bS_0 \cdot \bhL/|| \bL ||^2$,  which are truly conserved when ignoring radiative effects and spin-spin interactions. However, in the presence of these corrections $\Delta_{\text{eff L}}$, $\Delta (\bS_0 \cdot \bhL / || \bL ||^2  )$ and $\Delta_{0\text{L}}$ have larger variances than both $\Delta \cE_5^\text{NL}$ and $\Delta \cE_6^\text{NL}$.  This is illustrated in 
 Figure \ref{fig:norm_dist6d} where we show the distributions of $\Delta \cE_4^\text{NL}$, $\Delta \cE_5^\text{NL}$ and $\Delta \cE_6^\text{NL}$, normalized to their maximum possible value, for a binary black hole system with $m_1 = 0.4$ and maximal spin magnitudes. This choice of parameters corresponds to one of the largest variances for  
 these three principal components and is thus indicative of a black hole binary with some of the ``worst" statistics. We also show the normalized distributions for $\Delta_{0\text{L}}$ and $\Delta _{\text{eff L}}$, which would be identically zero if ignoring spin-spin and radiative effects. However, they are not when including those effects and instead we find that while $\Delta_{0\text{L}}$ and $\Delta _{\text{eff L}}$ are clustered around zero with a small variance of $\approx 5 \times 10^{-5}$ the quantities $\Delta \cE_5^\text{NL}$ and $\Delta \cE_6^\text{NL}$ both have a variance smaller by a factor of $\approx 10$, indicating that $\cE_5^\text{NL}$ and $\cE_6^\text{NL}$ are better conserved throughout the duration of the inspirals.

\begin{figure}
\includegraphics[width=\columnwidth]{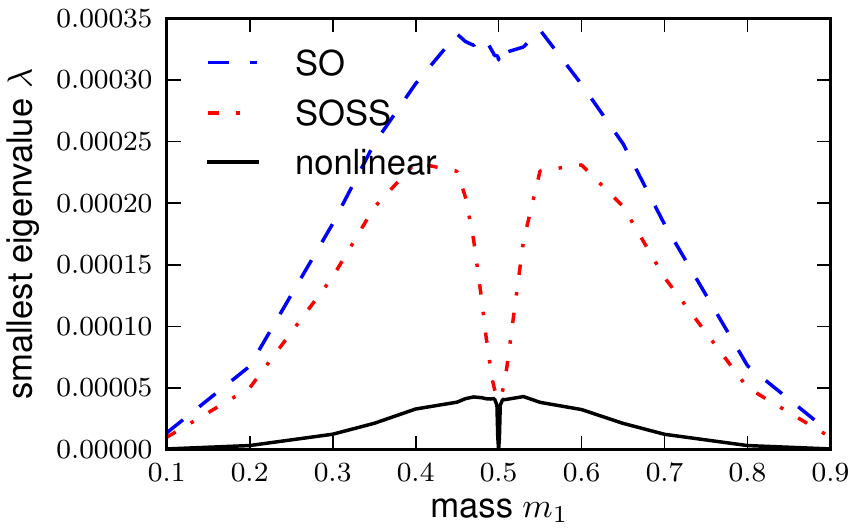}
\caption{Smallest eigenvalue as a function of one of the black hole masses (maximally spinning, with total mass set to $m_1+m_2=1$). 
Shown  are the purely spin-orbit (SO), the spin-orbit and spin-spin (SOSS),
  and the spin-orbit, spin-spin and nonlinear terms (nonlinear)
  cases. These correspond to eigenvalues $\lambda_2^{\text{SO}}, \lambda_3^{\text{SOSS}}$ and 
  $\lambda_6^{\text{NL}}$, respectively.  The equal-mass configuration has a near-maximum
  value for the SO case and a minimum in the other ones
  (with $\lambda_6^{\text{NL}}<10^{-12}$).}
\label{fig:m1_lam}
\end{figure}

Figure \ref{fig:m1_lam} shows the improvement obtained when including spin-spin variables in the PCA and when further 
including nonlinear spin-orbit ones. Shown is the smallest variance, for maximally spinning black holes, as a function 
of one of the black hole masses for the three PC analyses considered in this paper: 1) SO -- purely
spin-orbit terms, 2) SOSS -- spin-orbit and spin-spin terms, and 3)
Nonlinear -- spin-orbit, spin-spin and nonlinear terms. The variance becomes smaller with each 
term included in the PCA. For the purely
spin-orbit variables the smallest variance in the equal-mass case is
near maximum. In the other cases the equal mass configuration, in contrast, has the minimum variance 
with $\lambda_6^{\text{NL}}<10^{-12}$ for the PCA including the nonlinear variables.

%%%
\subsection{Binaries with equal masses and spin magnitudes}
%%%

Including the nonlinear spin-orbit variables into the principal component analysis reveals an interesting relation in the specific case of a binary black hole with equal masses and spin magnitudes undergoing a quasi-circular inspiral. It turns out that, as hinted above,  for this subset of parameter values $\lambda_6^\text{NL}$ is zero within numerical precision, which indicates that not only does $\Delta \cE_6^\text{NL}$ have a vanishing variance but, together with the observation that within such precision $\langle \Delta \cE_6 \rangle$ is also zero, that $\Delta \cE_6^\text{NL}$ is itself zero.  In other words,  $\cE_6^\text{NL}$ is  {\it truly conserved}, at least for the quasi-circular inspiral of binary black holes with equal masses and spin magnitudes. 

From the instantaneous approximation one can actually see that $\cE_6^\text{NL}$ is given by
\begin{equation}
	\cE_6^\text{NL} = \frac{2 \bhS_1 \cdot \bhS_2 + (\bhS_1 \cdot \bhL) (\bhS_2 \cdot \bhL) }{\sqrt{5} } \, . \label{E6_eq}
\end{equation}
and by using the PN equations of motion one can show that the time derivative of $\cE_6^\text{NL}$ vanishes.  The conservation of (\ref{E6_eq}) implies the angle between $\bhS_1$ and $\bhS_2$ is fixed at every time by the angles of the spin orientations with the orbital angular momentum.

%%%
\subsection{Predictions using principal components}
%%%

\begin{figure}
\includegraphics[width=\columnwidth]{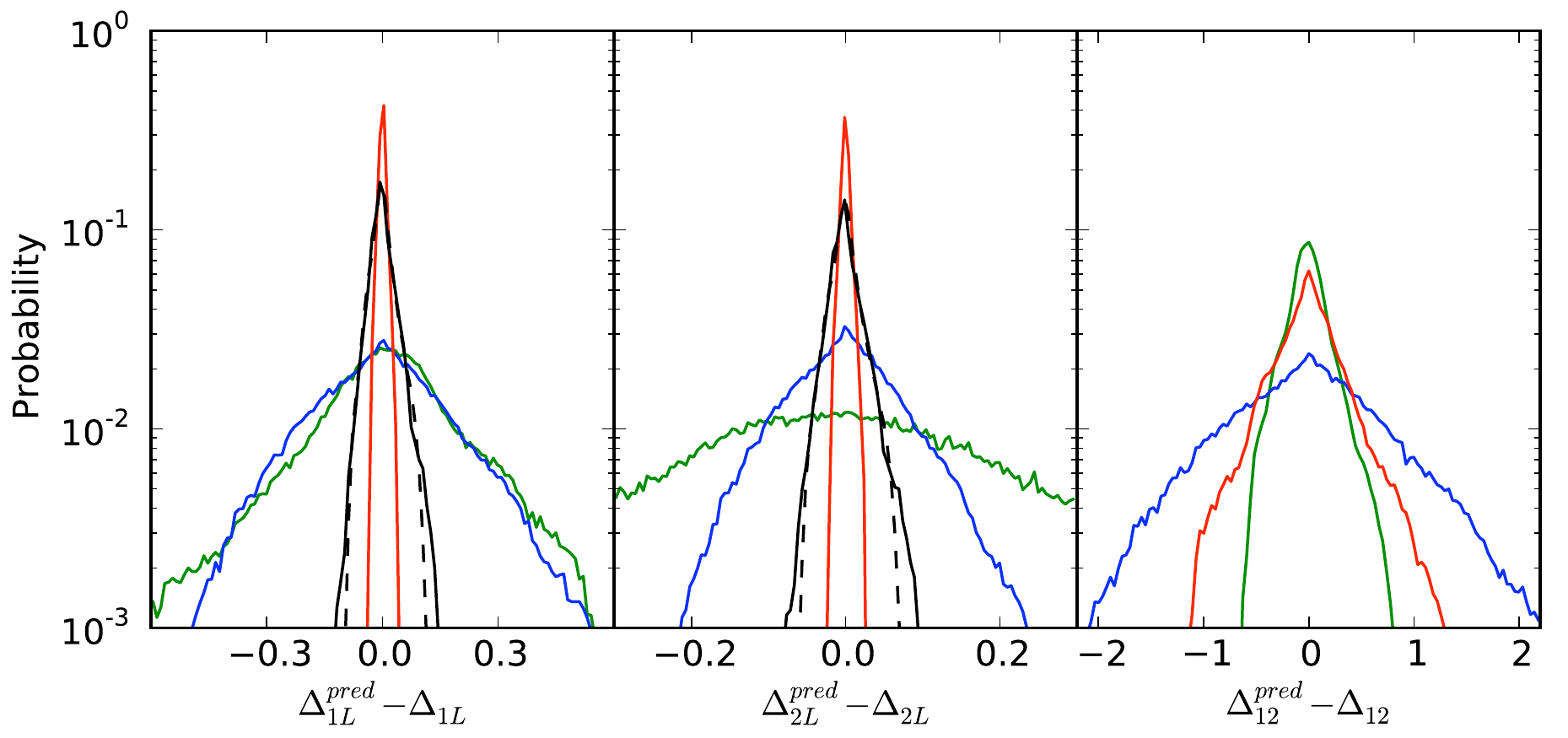}
\caption{Errors in the predictions of $\Delta_{1 \text{L}}$, $\Delta_{2\text{L}}$ and $\Delta_{12}$ for binary black holes with $m_1=0.4$ (total mass  $M=1$) and maximal spin magnitudes. The errors are defined as the predicted values (found by solving $\Delta \cE_4^\text{NL} = 0$ (green), $\Delta \cE_5^\text{NL} = 0$ (blue) and $\Delta \cE_6^\text{NL} = 0$ (red)) minus those from our numerical simulations. For comparison, we have also plotted the errors in these predictions when imposing $\Delta ( \bS_0 \cdot \bhL ) = 0$ (solid black) and $\Delta ( \bS_{\rm eff} \cdot \bhL ) = 0$ (dashed). Note that none of these last two conditions can be used to make a prediction for $\Delta_{12}$.}
\label{fig:prederr_6d}
\end{figure}

Given that $\cE_4^\text{NL}$, $\cE_5^\text{NL}$ and $\cE_6^\text{NL}$ are semi-conserved to varying degrees depending on the choice of binary black hole parameters it is tempting to use these quantities to predict the three relevant angles between the spin orientations and the orbital angular momentum. Figure \ref{fig:prederr_6d} shows the errors in predicting $\Delta_{1 \text{L}}$, $\Delta _{2 \text{L}}$ and $\Delta_{12}$ for one of the cases with largest 
variances. The errors are defined as the predicted values minus those obtained from our numerical simulations. In calculating the predicted ones, we assume that $\Delta \cE_{4,5,6}^\text{NL}$ are exactly zero and solve for the quantity of interest in terms of the other variables, the latter being replaced by their values from our numerical simulations.

Figure \ref{fig:prederr_6d} shows that the assumption $\Delta \cE_6^\text{NL}=0$ does quite well at predicting either $\Delta_{1\text{L}}$ or $\Delta_{2\text{L}}$ and gives better predictions than assuming that either $\Delta_{0\text{L}}$ or $\Delta_{\text{eff L}}$ are zero. In predicting $\Delta_{12}$ we see that assuming $\Delta \cE_4^\text{NL} = 0$ gives the best predictions, albeit not particularly good ones. Errors in approximating a principal component by exactly zero propagate in different ways into the scalar products when predicting the latter, according to their weights in the principal components (this is standard propagation of errors analysis). In particular, 
the large errors in predicting $\Delta_{12}$ are a consequence of the spin-spin interaction (and thus the corresponding weight in the principal component) being far weaker than spin-orbit corrections to the inspiral dynamics.

The predictions in Figure \ref{fig:prederr_6d} assume that one knows all variables but one with certainty. However, since we have three independent semi-conserved quantities, 
$$
	\Delta \cE_4^\text{NL} \approx 0 , ~~~ \Delta \cE_5^\text{NL} \approx 0 , ~~~ \Delta \cE_6^\text{NL} \approx 0 \, , 
$$
 and three angles uniquely describing the orientations of the spin and orbital angular momenta vectors, it is in principle possible to predict these angles based solely on their distributions at the initial time. Unfortunately, in general this gives very large errors. For example, in the maximal spinning and equal mass case the errors in this prediction for $\Delta_{1\text{L}}$, $\Delta_{2\text{L}}$ and $\Delta_{12}$ have, as probability distributions, standard deviations (not variances) of $0.32$, $0.32$ and $0.05$, respectively; and they are considerably larger for unequal masses. 

%%%%%%%%%%%%%%%%%%%%%%%%%%%%%
\section{General relativistic dynamics}
%%%%%%%%%%%%%%%%%%%%%%%%%%%%%

Since the PN approximation is highly accurate in the inspiral regime, 
one might expect that in the fully General Relativistic (GR) case the principal
components with smallest variances discussed in this paper could also have small variances when computed with full GR data. 
If so, a natural question would be whether that is still the case in the plunge regime. 

To have a preliminary sense of to what extent that might happen, we present three-dimensional numerical simulations of binary black holes using the full Einstein equations, starting around one and a half orbits before merger. 
Sampling the full parameter space of the binary black hole
problem through numerical evolutions of the full Einstein equations is
not possible. Even sampling a single configuration of masses and spin
magnitudes would be unfeasible if $40,\!000$ inspirals were
needed. However, the fact that in the PN approximation the
distributions of our principal components with smallest variances are so 
sharply peaked suggests that a small sample might be enough to
reproduce their main features. For example, in the equal mass and equal spin case one  
can coarsen the sampling within the PN approximation to $\sim 20$ spin
orientations and still reproduce the first significant digit of $\Delta \cE_3^{\text{SOSS}}$ when doing a PCA. 

We therefore present GR simulations of binary black hole configurations in an initial quasi-circular orbit with equal masses $m_1=m_2=0.5$ and spin magnitudes $\chi_1=\chi_2=0.6$, an initial separation of
$d=6.2\,M$ and 20 random initial spin directions $( \hat{\mathbf{S}}_1,\hat{\mathbf{S}}_2 )$. These configurations are evolved using the moving punctures
technique~\cite{Brownsville:PRL96,Goddard:PRL96} all the way through merger. The spin vectors 
are measured over time following the isolated horizon
approach~\cite{Dreyer02a,Ashtekar:2004cn,Herrmann:2007ex}. The statistical results shown below are insensitive from 
 removing some simulations from the analysis, hinting that such a small sample number might be sufficient to describe the salient features, just 
 as in the PN approximation.
\begin{figure}
\includegraphics[width=\columnwidth]{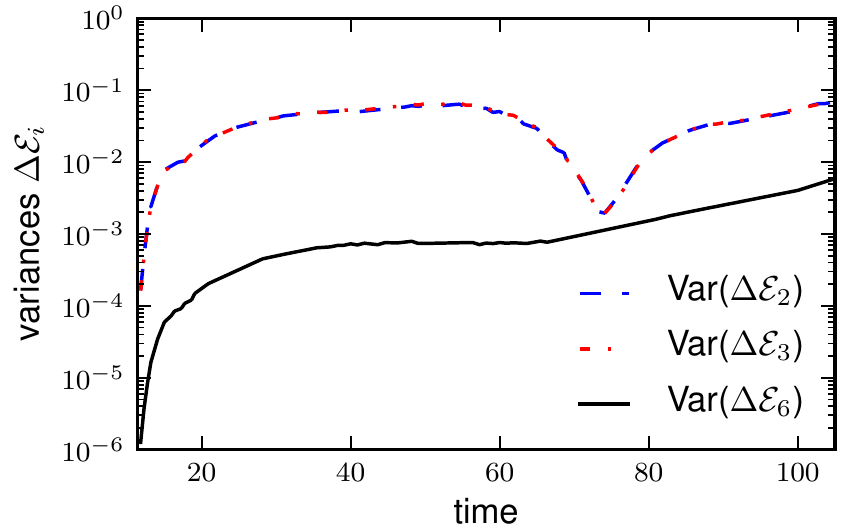}
\caption{Variances of $\Delta \cE_2^{\text{SO}}$, $\Delta \cE_3^{\text{SOSS}}$ and
   $\Delta \cE_6^{\text{NL}}$ as a function of time, using data from fully nonlinear
  GR simulations of binary black hole inspirals.}
\label{fig:gr}
\end{figure}

In Figure \ref{fig:gr} we show the variances for $\Delta \cE_2^{\text{SO}}$,
$\Delta \cE_3^{\text{SOSS}}$ and $\Delta \cE_6^{\text{NL}}$, as defined by our PN PCA analysis, Eqs.~(\ref{DE2_2d}), (\ref{lincomb_1}) and (\ref{E6_eq}), but where 
the data for the scalar products (\ref{delta1L})-(\ref{delta2L2L}) are taken from the fully relativistic simulations. The variances for $\Delta \cE_2^{\text{SO}}$ and
$\Delta \cE_3^{\text{SOSS}}$ appear indistinguishable from each other on the scale shown. The reason for this is that their difference is 
proportional to the spin-spin interaction, which is small (see Eqs.~(\ref{DE2_2d}), (\ref{lincomb_1}) and Figure \ref{fig:m1_lam}). All the variances are considerably larger than the PN corresponding ones but the main features of the PN 
analysis seem to hold. The variances of $\Delta \cE_2^{\text{SO}}$ and
$\Delta \cE_3^{\text{SOSS}}$ are larger than the variance of $\Delta \cE_6^{\text{NL}}$, as expected from our PCA analysis 
(see Fig.~\ref{fig:m1_lam}), and its conservation properties  (cf. the discussion near Eq.~(\ref{E6_eq})) at the PN level.  In particular, notice that $\Delta \cE_6 ^{\text{NL}}$ is approximately constant (after the initial transients have settled for $t \lsim 25M$) from about $25M$ to $75M$ after which it grows exponentially during the plunge phase from about $75M$ to the final time shown in the figure.
However, most importantly, the variance of our principal component $\Delta \cE_6^{\text{NL}}$ remains fairly small all the way until a common apparent horizon is found (which is just after the final time shown in Fig.~\ref{fig:gr}). 

The results of these simulations suggest that to some extent the
statistical structures identified in this paper within the PN approximation carry over to the inspiral and plunge 
regimes in the fully General Relativistic case.

\section{Conclusions}

In summary, we have identified statistical constraints in binary black
hole dynamics, from inspiral to merger, suggesting a dimensional
reduction of the parameter space when probabilistic
predictions are sufficient. We have also shown that these
probabilistic predictions can be surprisingly accurate using  
numerical simulations or, to a lesser degree, using the analytical
instantaneous approximation. 

When using purely spin-orbit variables for a principal component analysis we  have found that in the instantaneous approximation the 
PC with smallest variance, $\Delta\cE_2^{SO}$ (which is proportional to $\Delta(\mathbf{S}_0 \cdot \bhL)$), is a statistically
conserved quantity both in a statistical sense in terms of initial spin orientations but also as a function of time. We have also found the dominant component $\Delta\cE_1^{SO}$ describing the largest variation in the data. This demonstrates the ability of 
PCA to robustly identify conserved quantities as well as the most relevant features in the data. 

In our more general analysis we have identified three statistically uncorrelated semi-conserved quantities $\Delta \cE_{4,5,6}$ that do not correspond to any previously known expressions that we are aware of. These principal components are largely conserved when including spin-spin interactions and radiative corrections in the evolution to the extent that they are even reasonably conserved 
when using data from fully relativistic, three-dimensional simulations of binary black holes in the inspiral and plunge regime, just up to merger.

We expect these results to be useful in astrophysical
simulations of binary hole inspirals and in generating waveform templates
for gravitational wave data analysis. In particular, simplifying template generation may be realized with our PCA method by identifying those combinations of spin variables that are not only semi-conserved, and therefore effectively reduce the dimensionality of the parameter space (which may supplement the single-spin approximation used in \cite{Pan_etal:PRD69, Buonanno_etal:PRD70, Apostolatos_etal:PRD49}), but also those that encode the largest features in the data, and therefore are useful for efficiently sampling the parameter space thereby reducing the cost of constructing templates.

\section{Acknowledgments}

This work has been supported by NSF Grants PHY0801213 and PHY0908457 to the University of Maryland
and NVIDIA Corporation through a Professor Partnership award. The simulations were carried out at the Teragrid under allocation TG-PHY090080 with some of them using GPUs on the Lincoln cluster. We thank Alessandra Buonanno, Emanuele Berti, and Yi Pan for helpful discussions and suggestions. We also thank Five Guys Burgers and Fries, where some of this research was conducted.

\begin{appendix}\label{sec:appendix}
%%%%%%%%%%%%%%%%%%%%%%%%%%%%%%%
\section{Post-Newtonian equations of motion}
%%%%%%%%%%%%%%%%%%%%%%%%%%%%%%%
The Post-Newtonian equations used in this paper are those of Ref.~\cite{Buonanno:2002fy}, \cite{Buonanno:2002erratum}),

\begin{widetext}
\begin{eqnarray}
\dot{\omega}&=&\omega^2 \frac{96}{5} \eta (M\omega)^{5/3} \Bigg\{1-\frac{743+924 \eta}{336} (M\omega)^{2/3}\nonumber\\
            & &    -\left(\frac{1}{12} \sum_{i=1,2}\left(\chi_i \hat{\mathbf{L}}_n\cdot\hat{\mathbf{S}}_i (\frac{113 m_i^2}{M^2}+75 \eta)\right)-4 \pi\right) M\omega\nonumber\\
            & &    +\left( \frac{34103}{18144}+\frac{13661}{2016} \eta+\frac{59}{18} \eta^2 \right) (M\omega)^{4/3}-\frac{1}{48} \eta \chi_1 \chi_2 \left(247 (\hat{\mathbf{S}}_1\cdot\hat{\mathbf{S}}_2)-721 (\hat{\mathbf{L}}_n\cdot\hat{\mathbf{S}}_1) (\hat{\mathbf{L}}_n\cdot\hat{\mathbf{S}}_2)\right) (M\omega)^{4/3}\nonumber\\
            & &    -\frac{1}{672} (4159+15876 \eta) \pi (M\omega)^{5/3}+\Bigg(\left(\frac{16447322263}{139708800}-\frac{1712}{105} \gamma_E+\frac{16}{3} \pi^2\right)+(-\frac{273811877}{1088640}+\frac{451}{48} \pi^2-\frac{88}{3} \hat{\theta}\eta) \eta\nonumber\\
            & &    +\frac{541}{896} \eta^2-\frac{5605}{2592} \eta^3-\frac{856}{105} log(16 (M\omega)^{2/3})\Bigg) (M\omega)^2+(-\frac{4415}{4032}+\frac{358675}{6048} \eta+\frac{91495}{1512} \eta^2) \pi (M\omega)^{7/3} \Bigg\}\label{eq:domdt}\\
\dot{\mathbf{S}_i}&=&\mathbf{\Omega}_i \times \mathbf{S}_i\label{eq:dSidt}\\
\dot{\hat{\mathbf{L}}}_n&=&-\frac{(M\omega)^{1/3}}{\eta M^2} \frac{d\mathbf{S}}{dt}\label{eq:dLdt}
\end{eqnarray}
\end{widetext}
where $d\mathbf{S}/dt=d\mathbf{S}_1/dt+d\mathbf{S}_2/dt$,
$\gamma_E=0.577\ldots$ is Euler's constant, and
$\hat{\theta}=1039/4620$. The total mass is denoted by $M=m_1+m_2$ and
$\eta=m_1 m_2/M^2$ is the symmetric mass ratio. The magnitude of the
angular momentum can be computed via $||\mathbf{L}_n||=\eta
M^{5/3} \omega^{-1/3}$.

The evolution of the individual spin vectors $\mathbf{S}_i$ for the 2
black holes is described by a precession around $\mathbf{\Omega}_i$ with
\begin{widetext}
\begin{equation}
\mathbf{\Omega}_1=\frac{(M\omega)^2}{2M}\left(\eta (M\omega)^{-1/3} (4+3\frac{m_2}{m_1})\hat{\mathbf{L}}_n+1/M^2(\mathbf{S}_2-3(\mathbf{S}_2\cdot\hat{\mathbf{L}}_n)\,\hat{\mathbf{L}}_n)\right)\,,\label{eq:Omega}
\end{equation}
\end{widetext}
and $\mathbf{\Omega}_2$ is obtained by $1\leftrightarrow 2$.

\end{appendix}

\bibliographystyle{unsrt}
\bibliography{references}

\end{document}